\def\arxiv{true}
\documentclass[sigconf,screen]{acmart}

\usepackage{xspace}
\usepackage{amsmath}
\usepackage{amsfonts}
\usepackage{url}
\usepackage{marvosym}
\usepackage{ifthen}
\theoremstyle{plain}

\newcommand{\chatoDisplayMode}[1]{#1}

\definecolor{MyRed}{rgb}{0.6,0.0,0.0} 
\definecolor{MyBlack}{rgb}{0.1,0.1,0.1} 
\newcommand{\inred}[1]{{\color{MyRed}\sf\textbf{\textsc{#1}}}}
\newcommand{\frameit}[2]{
  \begin{center}
  {\color{MyRed}
  \framebox[.9\columnwidth][l]{
    \begin{minipage}{.85\columnwidth}
    \inred{#1}: {\sf\color{MyBlack}#2}
    \end{minipage}
  }\\
  }
  \end{center}
}

\newcommand{\note}[2][]{\chatoDisplayMode{\def\@tmpsig{#1}\frameit{{\Pointinghand} Note}{#2\ifx \@tmpsig \@empty \else \mbox{ --\em #1}\fi}}}
\newcommand{\todo}[2][]{\chatoDisplayMode{\def\@tmpsig{#1}\frameit{{\Writinghand} To-do}{#2\ifx \@tmpsig \@empty \else \mbox{ --\em #1}\fi}}}

\newcommand{\abbrevStyle}[1]{#1}

\newcommand{\ie}{\abbrevStyle{i.e.}\xspace}
\newcommand{\eg}{\abbrevStyle{e.g.}\xspace}
\newcommand{\cf}{\abbrevStyle{cf.}\xspace}

\newcommand{\etc}{\abbrevStyle{etc.}\xspace}

\newcommand{\xhdr}[1]{\vspace{1.7mm}\noindent{{\bf #1.}}}

\newcommand{\textcite}[1]{\citeauthor{#1} \shortcite{#1}}

\newcommand{\hide}[1]{}

\makeatletter
\newcommand{\iffont}[2]{\ifthenelse{\equal{\f@family}{#1}}{#2}{}}
\makeatother

\iffont{ptm}{
  \usepackage{mathptmx}

  \DeclareSymbolFont{greek}{OML}{cmm}{m}{n}
  \DeclareMathSymbol{\alpha}{\mathalpha}{greek}{"0B}
  \DeclareMathSymbol{\beta}{\mathalpha}{greek}{"0C}
  \DeclareMathSymbol{\gamma}{\mathalpha}{greek}{"0D}
  \DeclareMathSymbol{\delta}{\mathalpha}{greek}{"0E}
  \DeclareMathSymbol{\epsilon}{\mathalpha}{greek}{"0F}
  \DeclareMathSymbol{\zeta}{\mathalpha}{greek}{"10}
  \DeclareMathSymbol{\eta}{\mathalpha}{greek}{"11}
  \DeclareMathSymbol{\theta}{\mathalpha}{greek}{"12}
  \DeclareMathSymbol{\iota}{\mathalpha}{greek}{"13}
  \DeclareMathSymbol{\kappa}{\mathalpha}{greek}{"14}
  \DeclareMathSymbol{\lambda}{\mathalpha}{greek}{"15}
  \DeclareMathSymbol{\mu}{\mathalpha}{greek}{"16}
  \DeclareMathSymbol{\nu}{\mathalpha}{greek}{"17}
  \DeclareMathSymbol{\xi}{\mathalpha}{greek}{"18}
  \DeclareMathSymbol{\pi}{\mathalpha}{greek}{"19}
  \DeclareMathSymbol{\rho}{\mathalpha}{greek}{"1A}
  \DeclareMathSymbol{\sigma}{\mathalpha}{greek}{"1B}
  \DeclareMathSymbol{\tau}{\mathalpha}{greek}{"1C}
  \DeclareMathSymbol{\upsilon}{\mathalpha}{greek}{"1D}
  \DeclareMathSymbol{\phi}{\mathalpha}{greek}{"1E}
  \DeclareMathSymbol{\chi}{\mathalpha}{greek}{"1F}
  \DeclareMathSymbol{\psi}{\mathalpha}{greek}{"20}
  \DeclareMathSymbol{\omega}{\mathalpha}{greek}{"21}
  \DeclareMathSymbol{\varepsilon}{\mathalpha}{greek}{"22}
  \DeclareMathSymbol{\vartheta}{\mathalpha}{greek}{"23}
  \DeclareMathSymbol{\varpi}{\mathalpha}{greek}{"24}
  \DeclareMathSymbol{\varrho}{\mathalpha}{greek}{"25}
  \DeclareMathSymbol{\varsigma}{\mathalpha}{greek}{"26}
  \DeclareMathSymbol{\varphi}{\mathalpha}{greek}{"27}
  \DeclareSymbolFont{otone}{OT1}{cmr}{m}{n}
  \DeclareMathSymbol{\Gamma}{\mathalpha}{otone}{0}
  \DeclareMathSymbol{\Delta}{\mathalpha}{otone}{1}
  \DeclareMathSymbol{\Theta}{\mathalpha}{otone}{2}
  \DeclareMathSymbol{\Lambda}{\mathalpha}{otone}{3}
  \DeclareMathSymbol{\Xi}{\mathalpha}{otone}{4}
  \DeclareMathSymbol{\Pi}{\mathalpha}{otone}{5}
  \DeclareMathSymbol{\Sigma}{\mathalpha}{otone}{6}
  \DeclareMathSymbol{\Upsilon}{\mathalpha}{otone}{7}
  \DeclareMathSymbol{\Phi}{\mathalpha}{otone}{8}
  \DeclareMathSymbol{\Psi}{\mathalpha}{otone}{9}
  \DeclareMathSymbol{\Omega}{\mathalpha}{otone}{10}
  \DeclareSymbolFont{syms}{OML}{cmm}{m}{it}
  \DeclareMathSymbol{\partial}{\mathord}{syms}{"40}
  \DeclareMathAlphabet{\mathbold}{OML}{cmm}{b}{it}
  \DeclareSymbolFont{largesymbols}{OMX}{cmex}{m}{n}

}

\usepackage{graphicx} 
\usepackage{balance}
\usepackage{amsfonts,amsmath,amsthm}
\usepackage{bm}
\usepackage{booktabs} 
\usepackage{threeparttable}
\usepackage[htt]{hyphenat}
\usepackage{enumitem}
\usepackage{xspace}
\usepackage{multirow}
\usepackage{makecell}
\usepackage{hhline}
\usepackage{hyperref}
\usepackage{url}
\usepackage{colortbl}
\usepackage{subfig}
\usepackage[normalem]{ulem} 
\usepackage{algorithm}
\usepackage{appendix}   
\usepackage[noend]{algorithmic}
\usepackage{arydshln}

\setcitestyle{numbers,super&compress}

\newcommand{\ignore}[1]{}

\newcommand{\figcaption}[1]{\vspace*{-0mm}\caption{#1}\vspace*{-0mm}}

\newcommand{\qb}{\textsc{Quotebank}\xspace}

\ifx\arxiv\undefined
\else
\fi

\copyrightyear{2022}
\acmYear{2022}
\setcopyright{acmlicensed}

\ifx\arxiv\undefined
    \acmConference[SIGIR '22]{Proceedings of the 45th International ACM SIGIR Conference on Research and Development in Information Retrieval}{July 11--15, 2022}{Madrid, Spain}
    \acmBooktitle{Proceedings of the 45th International ACM SIGIR Conference on Research and Development in Information Retrieval (SIGIR '22), July 11--15, 2022, Madrid, Spain}
    \acmPrice{15.00}
    \acmDOI{10.1145/3477495.3531696}
    \acmISBN{978-1-4503-8732-3/22/07}
\else
  \acmConference[arXiv]{arXiv e-print}
  \acmBooktitle{}
  \acmPrice{}
  \acmISBN{}
  \acmDOI{10.1145/3477495.3531696}
\fi

\begin{document}
\fancyhead{}

\title{Quote~Erat~Demonstrandum: A~Web~Interface~for~Exploring~the~\qb Corpus}

\author{Vuk Vukovi\'c}
\affiliation{%
  \institution{EPFL}
  \country{}
}
\email{vuk.vukovic@epfl.ch}

\author{Akhil Arora}
\affiliation{%
  \institution{EPFL}
  \country{}
}
\email{akhil.arora@epfl.ch}

\author{Huan-Cheng Chang}
\affiliation{%
  \institution{EPFL}
  \country{}
}
\email{huan-cheng.chang@epfl.ch}

\author{Andreas Spitz}
\affiliation{%
  \institution{University of Konstanz}
  \country{}
}
\email{andreas.spitz@uni-konstanz.de}

\author{Ro\-bert West}
\affiliation{%
  \institution{EPFL}
  \country{}
}
\email{robert.west@epfl.ch}

\begin{abstract}
The use of attributed quotes is the most direct and least filtered pathway of information propagation in news. Consequently, quotes play a central role in the conception, reception, and analysis of news stories. Since quotes provide a more direct window into a speaker's mind than regular reporting, they are a valuable resource for journalists and researchers alike. While substantial research efforts have been devoted to methods for the automated extraction of quotes from news and their attribution to speakers, few comprehensive corpora of attributed quotes from contemporary sources are available to the public. Here, we present an adaptive web interface for searching \qb, a massive collection of quotes from the news, which we make available at \url{https://quotebank.dlab.tools}.
\end{abstract}



\maketitle


\section{Introduction}

Quotes of sources, politicians, athletes, or scientists play an important role in lending credibility to news articles~\cite{duncan2019don}. As news stories evolve, quotes are useful data for analyzing the spread of information through the news~\cite{DBLP:conf/www/NiculaeSZDL15}, for determining the source of news information~\cite{DBLP:journals/corr/abs-2104-09656}, or in fact checking and credibility assessment~\cite{DBLP:conf/www/PopatMSW18, DBLP:conf/cikm/CaoDGMT19}. Outside of journalistic applications, extracted quotes from the news are also valuable in social studies, for example through opinion mining from quotes~\cite{DBLP:conf/iat/BalahurSGPK09}.

The substance of quotes lies not just in \emph{what} is being said, but \emph{by whom} the quote is uttered since the attribution to a speaker provides context to the words. As a result, the automated extraction and attribution of quotes from document corpora has been the subject of ongoing research over the years \cite{DBLP:conf/emnlp/ParetiOKCK13, DBLP:conf/icwsm/PavlloP018, DBLP:conf/eacl/JurafskyCMF17, DBLP:conf/ranlp/PapayP19}. However, while methods for the extraction and attribution of quotes are developed continuously and benefit from recent advances in natural language processing, few of the resulting resources are available to end-users. To address this need, we report on the development of a user interface for searching \qb~\cite{DBLP:conf/wsdm/VaucherSC021}, a massive corpus of quotes that we extracted from a decade of English news.

\textbf{Contributions.} 
We provide an interface for (faceted) search in \qb, a Web-scale database of quotes from a decade of English news articles. Our tool is accessible as a website that is geared towards end-users who would otherwise be unable to use this resource, and it provides near-realtime query performance that enables an interactive exploration of the \qb corpus.


\section{Related Work}

Prior work that is related to \qb is quite sparse and can be grouped into two categories: \emph{corpora of quotes} and \emph{system demonstrations} that make use of or recommend quotes.

\xhdr{Quote corpora} 
Available corpora include the PolNeAR corpus~\cite{DBLP:conf/lrec/NewellMR18}, the Penn Attribution Relation Corpus~\cite{DBLP:phd/ethos/Pareti15}, the Speech, Thought, and Writing Presentation corpus~\cite{DBLP:journals/lalc/Hardy07}, the Rich Quotation Annotations corpus~\cite{DBLP:conf/lrec/PapayP20}, and the DirectQuote dataset~\cite{DBLP:journals/corr/abs-2110-07827}. In contrast to \qb, 
all of the above corpora are available only as NLP resources, not as searchable repositories that can be used by laypeople. Furthermore, they are substantially smaller in comparison to \qb, which exceeds the amount of contained quotes in these corpora by several orders of magnitude.

\xhdr{System demonstrations} 
To the best of our knowledge, there are no other search interfaces for large-scale corpora of quotes from the news domain. Repositories of quotes that can be accessed and searched by laypeople remain limited to manually curated repositories of mostly historical quotes, such as WikiQuote.\footnote{\url{https://www.wikiquote.org/}} System demonstrations in the information retrieval community tend to only consider quotes indirectly and often in the wider contexts of credibility assessment, such as CredEye~\cite{DBLP:conf/www/PopatMSW18}, or fact checking for news such as BeLink~\cite{DBLP:conf/cikm/CaoDGMT19} and the work of Miranda et al.~\cite{DBLP:conf/www/MirandaNMVSGMM19}.

\begin{figure*}[t]
    \centering
    \includegraphics[width=0.9\linewidth]{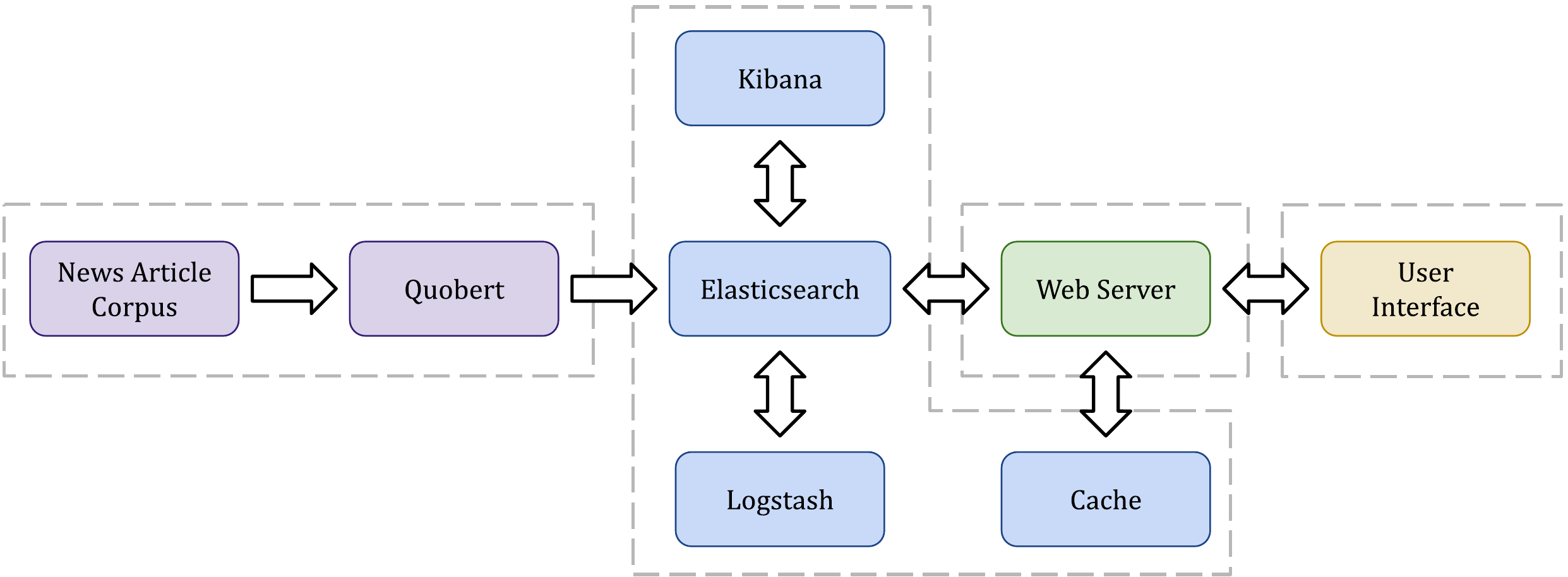}
    \figcaption{Schematic overview of the \qb system architecture. During pre-processing, speaker candidates and quotes are annotated in the news article corpus, and quotes are attributed to speakers with Quobert~\cite{DBLP:conf/wsdm/VaucherSC021}. The attributed quote data is augmented by linking speakers to Wikidata and subsequently loaded into Elasticsearch, which is used for indexing and retrieval. Kibana and Logstash are used internally for monitoring and managing indices. The web server exposes the query API to handle user queries and directly interfaces with the cache to reduce the impact of complex queries on system load. The JavaScript UI translates user queries to request to the web server (for details on the UI, see Figure~\ref{fig:ui}). }
    \label{fig:arch}
\end{figure*}

The likely most closely related approach to ours is by MacLaughlin et al., who propose a system for quote recommendation based on context by using a BERT architecture~\cite{DBLP:conf/icwsm/MacLaughlinCAR21}. However, the employed QUOTUS data set~\cite{DBLP:conf/www/NiculaeSZDL15} is relatively small and the tool is neither available nor suitable for the exploration of a Web-scale news corpus that would be useful to an end-user.


\section{Quote Corpus}
\label{sec:corpus}
\qb is a dataset of 235 million unique, speaker-attributed quotes that were extracted from 196 million English news articles (127 million of them containing quotes) published between September 2008 and April 2020~\cite{DBLP:conf/wsdm/VaucherSC021}. The data was extracted with the BERT-based architecture Quobert, which utilizes Quootstrap~\cite{DBLP:conf/icwsm/PavlloP018} for the extraction of training data. We make two stages of the data available in our search interface: \emph{article-level} and \emph{quote-level}.

\xhdr{Article-level data} 
In the article-level version of the data, articles are the central unit. The data contains individual quote annotations for all articles in the news data set, each attributed with a ranked list of the most likely speaker candidates in the article (or \emph{no speaker} if no suitable candidate could be attributed by Quobert). Additionally, the data contains context windows surrounding each of the quote mentions in the news article text.

\xhdr{Quote-level data} 
The quote-level data contains an aggregated view of the quotes across all articles. To generate this stage of the data, all individual occurrences of quotes are first canonicalized and quotes with matching canonical form are then aggregated into a single data point. Speaker candidates for these quotes are merged by weighted consensus, i.e., by summing over the local candidate probabilities in all individual occurrences to derive the most likely global speaker for a given quote.

\xhdr{Speaker data} To improve the querying capability of \qb and support faceted search, we further enrich the quote corpus with speaker information extracted from Wikidata~\cite{DBLP:journals/cacm/VrandecicK14}, which is one of the largest publicly available knowledge graphs (about 97M entities). Specifically, we extract and add data concerning the occupation, nationality, and gender of the speakers in \qb.

A full JSON dump of the raw data prior to the addition of speaker data from Wikidata is also available for download.\footnote{\url{https://doi.org/10.5281/zenodo.4277311}} For details on the generation, we refer the reader to Vaucher et al.~\cite{DBLP:conf/wsdm/VaucherSC021}.


\section{The \qb System}

The complete architecture of the \qb system is shown in Figure~\ref{fig:arch}. 
It consists of three core components: (1) a \emph{database system} housing the core storage and querying capabilities; (2) a \emph{web server} providing a layer of abstraction and API on top of the the database system; and (3) a \emph{user interface} responsible for delivering the results to the user through an interactive and flexible visualization.
In the following, we provide a description of these components and highlight their key functionality.

\subsection{Database system}
The \qb database is built using Elasticsearch \cite{elastic_search}, which is one of the most popular distributed, scalable, and open-source search and analytics engines supporting full-text indexing and querying, thereby serving our primary goal of efficiently searching through terabytes of quotes and news content.

Our database consists of three indices: (1) article, (2) quote, and (3) speaker, which store and index the article-level, quote-level, and speaker data, respectively. Na\"ively indexing the \qb corpus using Elasticsearch would require more than $2$TB of disk space, while the storage footprint of our optimized database is $4$~times~smaller and consumes only about $500$GB. We employ the following fundamental tried-and-true design decisions and optimizations to reduce the storage footprint and improve the query efficiency.
\begin{itemize}[leftmargin=*,noitemsep]
    \item \textbf{Database normalization}. We follow standard principles such as removing data redundancy, unless redundancy entails significant query speed-ups.
    \item \textbf{Choice of data types}. We use data types with minimal storage requirements, such as the \verb|integer| type for fields supporting range queries (\eg, number of occurrences of a quote), or \verb|keyword| type for fields used for creating filters (\eg, speaker nationality).
    \item \textbf{Querying-indexing trade-off}. We push the complexity to the indexing phase by aggregating all text type fields into a single field, which results in faster querying (at the cost of slower indexing) in comparison to searching multiple fields at query time.
\end{itemize}

\begin{figure*}[t]
    \centering
    \subfloat[]{
        \includegraphics[width=0.62\linewidth]{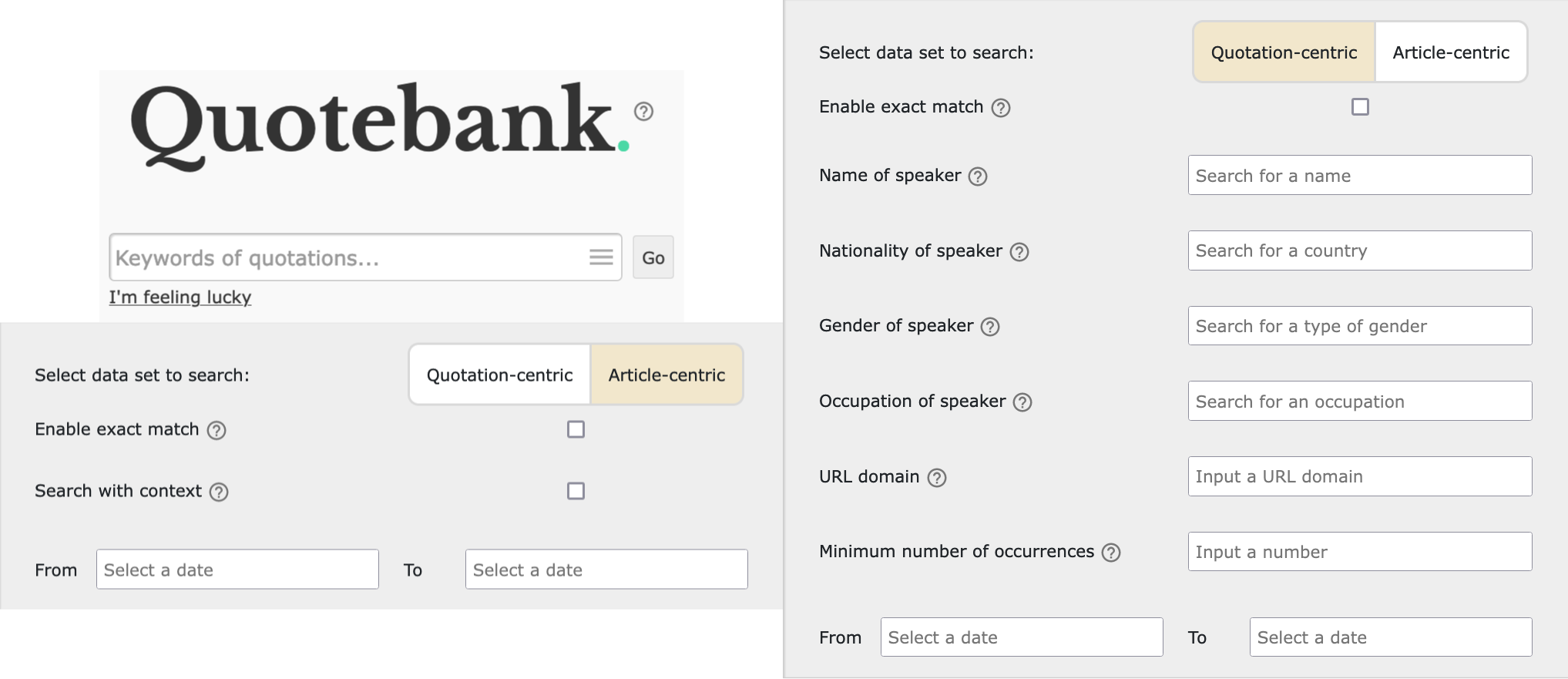}
        \label{fig:search_panel}
    }
    \subfloat[]{
        \includegraphics[width=0.39\linewidth]{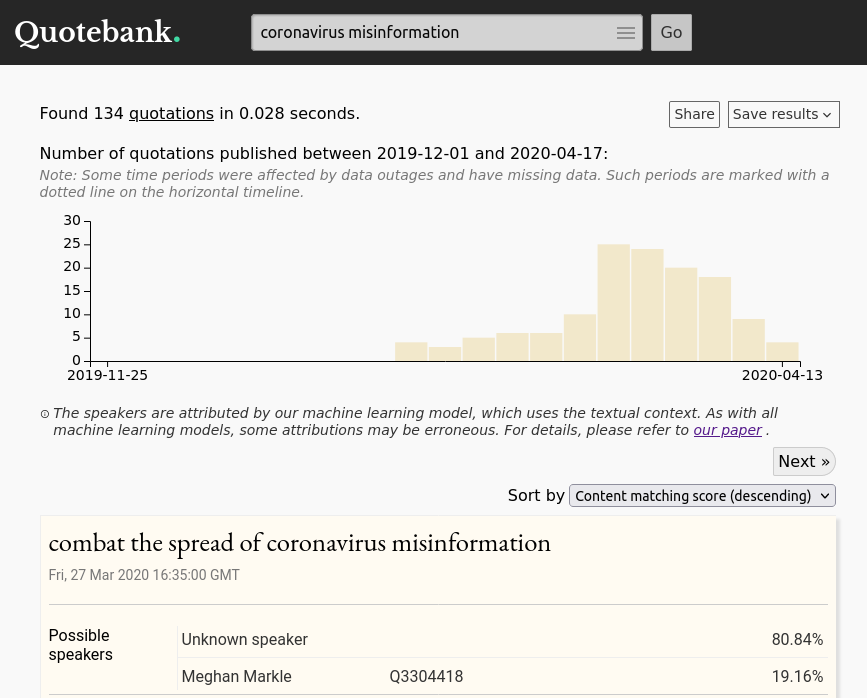}
        \label{fig:serp}
    }
    \figcaption{The \qb user interface. (a) The search panel supports quote-level and the article-level searches, which can be faceted by speaker attributes. (b) The search engine result page (SERP) displays retrieved quotes in the selected time window alongside speaker candidates and the URLs of articles from which the quotes are sourced. A histogram shows the distribution of quotes in the result over time.}
    \label{fig:ui}
\end{figure*}

The database system also houses \emph{Kibana} and \emph{Logstash}, which are primarily used for internal monitoring purposes. While Kibana is used for monitoring the status of indices and analyzing database performance statistics, Logstash handles storage management of the logs produced by Elasticsearch.

\subsection{Web server}
The web server provides a layer of abstraction on top of the database system and exposes the API endpoints that enable the communication between the user-interface and the database. It is responsible for composing, validating, and submitting user queries to the database as well as parsing and returning the retrieved result from the database to the user. In a nutshell, the web server prevents the users from communicating directly and in an unrestrained manner with the database, thereby providing an added level of security.

\subsection{User interface}
The user interface (\cf Figure~\ref{fig:ui}) is an adaptive web-based search platform built with React.js,\footnote{\url{https://reactjs.org/}} which allows users to interactively explore the \qb corpus. It consists of two main views: (1) the search panel, and (2) the search engine result page.

\xhdr{Search panel} Figure~\ref{fig:search_panel} portrays the interface that is exposed to the users for querying the \qb corpus. In the following, we describe the key features of the search panel.

\subsubsection{Corpus type} While the system supports queries on the quote-level data (called quotation-centric in Figure~\ref{fig:search_panel}) by default, it also provides the user with the option to query the article-level data.

\subsubsection{Query type} The most straight-forward way to search the \qb corpus is via \emph{text query}, which supports \emph{fuzzy matching} by default. However, the user may also choose to utilize \emph{exact matching} of the query text by enabling the checkbox option \verb|'Enable exact| \verb|match'|. For the article-level data, which supports only \emph{text queries}, users are able to query both the quotes and the context in which they occur in the original articles (\verb|'Search with context'| checkbox). To enable search faceting, the interface supports a multitude of search filters for querying quote-level data (\cf Figure~\ref{fig:search_panel}), such as speaker name or nationality, minimum number of quote occurrences, \etc, thereby providing additional refinements over and above the text query. By default, the time window is set to the full date range of the \qb corpus corresponding to September 1, 2008 and April 17, 2020, but the user may adapt this window by using the \verb|From| and \verb|To| fields, respectively. All other filter-related fields are not initialized with default values.

\subsubsection{Auto-complete} To assist users in applying search filters and increase search precision, auto-complete is implemented for each speaker-related field. Similar to most auto-complete implementations, users are only required to input the initial characters of a speaker's name in the corresponding field and may choose from the provided suggestions. The auto-complete function also provides a short description of the speaker (extracted from Wikidata), which is displayed alongside each suggestion to help users disambiguate and choose from the list of potential suggestions.

\xhdr{Search engine result page (SERP)} Figure~\ref{fig:serp} portrays the SERP, which consists of three primary components: a summary section, a histogram of quote occurrences, and the main result blocks.

\subsubsection{Summary section} This section displays basic information about the retrieved result, such as the number of quotes or articles that were deemed relevant to the query, and the query time in seconds. To reduce the load on the client, the total number of results returned from the web server is capped at 1000, corresponding to the $1000$~most-relevant results based on their content matching score with the user query. Additionally, this section provides a \verb|'Share'| button, which generates a shareable \emph{permalink} to the user's query and copies it to the clipboard. A \verb|'Save results'| button enables the user to download the results either as a \emph{JSON} or text file.

\subsubsection{Histogram} The histogram, implemented in D3.js,\footnote{\url{https://d3js.org}} displays the distribution of quotes or articles that match the user query in the specified time window. The histogram bin boundaries are automatically adjusted to improve its aesthetics while simultaneously keeping the resource utilization on the client side low. The main purpose of the histogram is to allow the user to effectively visualize the trend portrayed by results relevant to her query. While the returned results are capped at $1000$, note that  the histogram is based on all quotes or articles that match the user's query. Thus, the counts displayed in the histogram may exceed the number of returned documents.

\subsubsection{Result block} The returned results are displayed in blocks, where each block contains either a single relevant quote or a relevant article that contains at least one matching quote. The font and layout of the result block is designed to simulate a real quote as one might encounter it in a newspaper. The URLs in each block are clickable links to the news articles in which the quote was originally published. For quote blocks, all possible speakers of a quote are annotated with their unique Wikidata identifiers, which are also clickable links pointing to the Wikidata page of the speaker. Long result blocks are collapsed to a fixed size, allowing the user to effectively browse the result summary and only expand specific blocks if desired. In the same vein, we also implement pagination by breaking down the result blocks into pages of at most $10$ blocks each. The buttons \verb|Prev| and \verb|Next| can be used to navigate the pages. Lastly, we also enable the user to re-rank the returned results based on several pre-defined sorting criteria.

\section{Demonstration}

As part of the demonstration, we encourage the reader to engage with three visual scenarios crafted to highlight key features of the interface, namely the exploration of (1) quote-level and (2) article-level data, and (3) a free-roam exploration on a device of the user's choice (\eg, tablet, phone, or laptop). These scenarios were designed with support from an EPFL journalist, who also helped in improving our system's usability.

\subsection{Quote-level exploration}
In this scenario, the users may explore the \qb corpus to identify all the quotes from \emph{Donald Trump} containing the text ``great again'' and that appeared in at least $500$ distinct articles. They are further encouraged to re-rank the results in reverse chronological order and share the results with their colleagues by sending the \emph{permalink} via e-mail or messenger service.

\xhdr{User interaction} This scenario provides a demonstration of the \verb|'Quotation-centric'| (\cf Figure~\ref{fig:search_panel}) search panel. In addition to simply posing the text query ``great again'', the user has the option of applying multiple filters. Specifically, the user should set the \verb|'Minimum number of occurrences'| to $500$, and leverage the \emph{auto-complete} functionality to set the input for \verb|'Name of speaker'| to ``Donald Trump''. Lastly, the user may select the pre-defined filter \verb|'Date (descending)'| to re-rank the results in reverse chronological order. A permalink of the query is obtained by clicking on the \verb|'Share'| button on the results page. The retrieved results for this scenario can be accessed at this \href{https://quotebank.dlab.tools/search?target=quotation&text=great+again&speaker=Q22686&from_date=2008-09-01&to_date=2020-04-17&num_occurrences=500}{quote-level permalink}.

\subsection{Article-level exploration}
In this scenario, the user is encouraged to explore the \qb corpus to identify all articles published on May 19, 2018 that contain the text ``gdpr'' in either the quotes or in their surrounding context. The user should also download the results as a text file and view its contents in a text editor.

\xhdr{User interaction} This scenario provides a demonstration of the \verb|'Article-centric'| (\cf Figure~\ref{fig:search_panel}) search panel. In addition to posing the text query ``gdpr'', the user may vary the time window by restricting it to a particular day, \ie, May 19, 2018. Moreover, the user analyzes the differences in the retrieved results by toggling the \verb|'Search with context'| checkbox. Lastly, to facilitate later exploration and analyses, the user may download the results corresponding to her query as a text file  by clicking on the \verb|'Save results'| button on the results page. The retrieved results for this scenario can also be accessed at this \href{https://quotebank.dlab.tools/search?target=article&text=gdpr&from_date=2018-05-19&to_date=2018-05-19&with_context=true}{article-level permalink}.

\subsection{Free-roam exploration}
While we provide a few example use-cases (described below) to bootstrap the exploration in this scenario, the reader is encouraged to conduct a free-roam exploration of \qb. 
\begin{itemize}[leftmargin=*, noitemsep]
\item \textbf{UC 1:} Using \qb on their phone, users search quotes from female tennis players of Switzerland that appear in at least 100 different articles and share the results via messenger.
\item \textbf{UC 2:} Using \qb on their tablet, users search quotes related to ``science'' by female journalists that appear in at least 100 different articles and share the results via e-mail.
\item \textbf{UC 3:} Using \verb|'Enable exact match'| to search for a quote that the user remembers. For instance, the famous quote: ``You have to dream before your dreams can come true'' to recall the name of the speaker or identify its popularity over the past decade.
\end{itemize}

\section{Conclusion}
In this paper, we introduced a search interface that makes the \qb dataset more accessible to end users who lack the computational background to work directly with the raw data. To support faceted search based on speaker attributes, we further enriched the quote corpus with information from Wikidata. Our intention is to enable the general public to explore the \qb data, and we are looking forward to seeing the findings and insights that are gained in the exploration of the data by journalists, social scientist, and laypeople. 

\xhdr{Future work} 
In the future, we plan on analyzing the search logs and investigate users' search patterns to determine exploration strategies and use-cases that can aid us in further refining and augmenting the \qb corpus. We are also working on disambiguation techniques for speaker candidates during quote attribution and are preparing a stand-alone end-to-end pipeline for attributed quote extraction.

\xhdr{Acknowledgments} 
We would like to thank Tanya Petersen for testing the interface and sharing her expert-user insights. We are also grateful to the members of EPFL DLAB and the students enrolled in the 2021 edition of the Applied Data Analysis (ADA) course for their feedback. This project was partly funded by the Swiss National Science Foundation (grant 200021\_185043), the European Union (TAILOR, grant 952215), and the Microsoft Swiss Joint Research Center. We also gratefully acknowledge generous gifts from Facebook and Google.

\bibliographystyle{ACM-Reference-Format}
\balance
\bibliography{qed.bib}

\end{document}
\endinput